\documentclass{article}
\usepackage{arxiv}
\usepackage[utf8]{inputenc}
\usepackage[T1]{fontenc}
\usepackage{hyperref}
\usepackage{url}
\usepackage{booktabs}
\usepackage{amsmath,amssymb,amsthm}
\usepackage{bm}
\usepackage{graphicx}
\graphicspath{{figures/}}
\usepackage{xcolor}
\usepackage[numbers,sort&compress]{natbib}
\usepackage{enumitem}

\newtheorem{theorem}{Theorem}[section]
\newtheorem{lemma}[theorem]{Lemma}
\newtheorem{proposition}[theorem]{Proposition}

\theoremstyle{definition}
\newtheorem{example}[theorem]{Example}
\newtheorem{assumption}{Assumption}
\theoremstyle{remark}

\newcommand{\Cov}{\mathrm{Cov}}
\newcommand{\R}{\mathbb{R}}
\newcommand{\Vc}{\mathcal{V}}
\newcommand{\Lc}{\mathcal{L}}
\newcommand{\PoM}{\mathrm{Dist}}
\newcommand{\sd}{\mathrm{sd}}

\title{Strategic Opinion Manipulation in Multiplex Networks}

\author{Raman Ebrahimi\\
Department of Electrical and Computer Engineering\\
University of California, San Diego\\
\texttt{raman@ucsd.edu}
\And
Massimo Franceschetti\\
Department of Electrical and Computer Engineering\\
University of California, San Diego\\
\texttt{massimo@ucsd.edu} 
}

\begin{document}
\maketitle

\begin{abstract}
Models of opinion dynamics on networks, and DeGroot averaging in particular, provide a framework to study how a network of agents aggregates dispersed opinions into a consensus. However, existing models assume that agents truthfully report their opinions, and do not account for environments in which a platform aggregates agents' reports across \emph{multiple} networks (e.g., several online and offline channels), and agents can strategically report different opinions on each. In this paper, we propose a model of strategic opinion manipulation on multiplex networks, in which a platform merges $L$ network layers with attention weights, and each agent, holding a private opinion, chooses (potentially different) reports on each layer, at a layer-specific misreporting cost, so as to pull the merged consensus toward their own opinion. We show that this game has a unique Nash equilibrium in closed form, that the resulting consensus is the truthful consensus under a centrality \emph{tilted} toward a \emph{manipulability index} of each agent, and that the resulting distortion is the (centrality-weighted) covariance of agents' manipulability and opinions. We further show that agents' reports on every layer are more extreme than their opinions, so that the polarization observed by the platform overestimates the true polarization. Notably, we highlight that merging layers is a double-edged sword: as manipulability depends on the \emph{square} of an agent's influence on each layer, spreading attention across layers dilutes manipulation, while heterogeneous misreporting costs across layers and shifts in agents' centralities can amplify it. Finally, we characterize the platform's optimal choice of attention weights, in closed form when the layers share a stationary distribution, and through an exact gradient and a marginal test otherwise. Together, our findings shed light on when aggregating opinions across networks is (not) robust to strategic manipulation, and point out potential interventions (e.g., linking identities across platforms) to alleviate it.
\end{abstract}


\section{Introduction}\label{sec:intro}

Networks provide a powerful framework for understanding how opinions form, spread, and aggregate in a society. In particular, models of opinion dynamics such as DeGroot averaging~\cite{degroot1974reaching} and its extensions~\cite{golub2010naive,demarzo2003persuasion,acemoglu2013opinion,ghaderi2014opinion} have been used to study when a network of agents reaches a consensus, and whether that consensus aggregates the agents' dispersed information well. In these models, the long-run consensus is a weighted average of agents' initial opinions, with weights given by the agents' (eigenvector) centralities in the network of influence.

Although existing works capture the network over which opinions are exchanged, they generally assume that the opinion an agent feeds into the averaging process is the opinion they in fact hold. This assumption is questionable in environments where the outcome of the aggregation matters to the agents, and where the aggregation is carried out by a platform over \emph{multiple} networks. For instance, a city council may read public comments on a zoning proposal from an online portal, a town-hall meeting, and a neighborhood mailing list, and treat the three as one body of public opinion; a platform may score the sentiment around a product across a forum, a short-video feed, and a review site, and report a single number. In each case, the agents being aggregated know that they are being aggregated, and they need not say the same thing everywhere: a resident can be measured on the portal and strident at the town hall, and a firm's advocates can be muted under their own names and loud on an anonymous board. Misreporting one's opinion is not free (it costs reputation, effort, or the discomfort of saying something one does not believe), but this cost differs by channel, and so does the influence a statement carries on each channel.

Motivated by this, in this paper, we propose a model of \emph{strategic opinion manipulation on multiplex networks}. We build on the merged-layer coordination games of \cite{shiu2026coordination}, where a platform merges the $L$ layers of a multiplex network (each capturing one modality of interaction) using attention weights $\alpha$, and the resulting opinion dynamics converge to a consensus. We extend this model by letting each agent hold a private opinion, and choose the opinion they report on each layer strategically: an agent wants the merged consensus to be close to their own opinion, but pays a (layer-specific) quadratic cost for reporting an opinion different from the one they hold. We then ask what consensus emerges, what the platform observes, and, in particular, whether merging more layers makes the consensus more or less robust to manipulation.

\vspace{0.05in}
\subsubsection*{Paper overview and contributions} We consider $N$ agents who interact over $L$ layers, with row-stochastic influence matrices $W^{(l)}$ that the platform merges into $C(\alpha)=\sum_l\alpha_lW^{(l)}$. Our first step is to establish how the merged dynamics aggregate agents' \emph{per-layer} reports (Lemma~\ref{lem:aggregation}): the consensus is a weighted sum of reports, in which agent $i$'s report on layer $l$ carries a \emph{leverage} $\omega_{i,l}$ equal to the share of the network's long-run attention that reaches $i$ through layer $l$; an agent's leverages sum, across layers, to their centrality $\pi_i$ in the merged network. We then show that the resulting game admits a unique Nash equilibrium, in closed form (Theorem~\ref{thm:equilibrium}). Two quantities organize the analysis: the \emph{manipulability index} $K_i=\sum_l\omega_{i,l}^2/c_{i,l}$ of each agent (their squared leverage on each layer, discounted by their misreporting cost on that layer), and the resulting \emph{tilted centrality} $(\pi_i+K_i)/(1+\sum_jK_j)$ under which the equilibrium consensus is the truthful consensus. The distortion of the consensus relative to its truthful counterpart is then given by the centrality-weighted covariance of agents' manipulability and opinions (Theorem~\ref{thm:distortion}); in other words, strategic manipulation damages the consensus only when the agents with leverage are also the agents with extreme opinions.

We next consider what the platform observes. We show that every agent's report, on every layer, is a stretched version of their opinion, away from the consensus (Proposition~\ref{prop:posturing}), so that the polarization measured by the platform overestimates the true polarization; we also show that the ratio of an agent's distances from the consensus on two layers identifies their relative leverage on those layers, regardless of their opinion (Proposition~\ref{prop:signature}). We then turn to the paper's central question, of whether merging layers helps. Here, we identify a \emph{double-edged sword}: on the one hand, because leverage enters the manipulability index as a square, spreading the platform's attention across layers dilutes manipulation, and adding \emph{any} new layer at a small enough weight reduces distortion (Theorem~\ref{thm:dilution}); on the other hand, merging can hurt through two channels, namely, heterogeneity of misreporting costs across layers, which we bound by a Cauchy-Schwarz gap that linking agents' identities across layers closes (Theorem~\ref{thm:linking}), and \emph{re-centralization}, whereby the new layer shifts centrality toward agents with extreme opinions (Example~\ref{ex:recentralization}). Finally, we consider the platform's choice of attention weights $\alpha$. When the layers share a stationary distribution, we characterize the distortion-minimizing weights in closed form (Theorem~\ref{thm:design}); for general layers, we derive an exact gradient of distortion (through the derivative of the stationary distribution of the merged network), a marginal test for whether a new layer helps, and an exact algorithm for the two-layer case (Theorems~\ref{thm:gradient} and~\ref{thm:algebraic}). Each result is followed by an intuitive interpretation of its implications; all proofs are in Appendix~\ref{app:proofs}.

\vspace{0.05in}
\subsubsection*{A key takeaway} We highlight one recurring intuition gained from our analyses: when agents can misreport on several layers, the square in the manipulability index means that an agent's ability to move the consensus is governed by the \emph{concentration} of the platform's attention, and not only by its level. Intuitively, an agent who misreports on $L$ channels pays $L$ separate (convex) costs, while the influence bought on each channel is scaled by the platform's attention to it; merging layers therefore, by default, fragments agents' leverage. Merging becomes harmful only when it lets agents concentrate their manipulation where lying is cheap, or when it concentrates centrality on the agents whose opinions are far from the consensus. These findings can inform platform design: a platform can assess, before merging a new channel, whether the re-centralization it induces outweighs the dilution it brings (Theorem~\ref{thm:gradient}), and interventions that link an agent's identity across channels remove the cost-heterogeneity effect (Theorem~\ref{thm:linking}).

The remainder of this paper is organized as follows. We review the most closely related work in Section~\ref{sec:related}. Section~\ref{sec:model} introduces our model. We analyze the equilibrium and the resulting distortion in Section~\ref{sec:equilibrium}, and what the platform observes in Section~\ref{sec:observables}. Section~\ref{sec:multiplexing} studies whether merging layers helps, and Section~\ref{sec:design} the platform's choice of attention weights. We illustrate our findings numerically in Section~\ref{sec:numerics}, and conclude in Section~\ref{sec:conclusion}.

\section{Related Work}\label{sec:related}

Our work is at the intersection of three lines of literature: (i) opinion dynamics and social learning on networks, (ii) games on multiplex and multilayer networks, and (iii) strategic misreporting and manipulation-resistant aggregation.

\emph{Opinion dynamics on networks.} Our aggregation rule is DeGroot averaging~\cite{degroot1974reaching}, whose long-run consensus weights agents by the stationary distribution of the influence matrix; \cite{golub2010naive} characterize when this consensus aggregates dispersed information well, and \cite{demarzo2003persuasion} interpret the same weights as a persuasion bias. Our work is most closely related to game-theoretic treatments of these dynamics. Specifically, \cite{bindel2015forming} view Friedkin-Johnsen dynamics as the equilibrium of a game in which each agent trades off the distance from an internal opinion against disagreement with their neighbors, and bound the resulting price of anarchy; \cite{ghaderi2014opinion,acemoglu2013opinion} study the same dynamics with stubborn agents, and \cite{yildiz2013binary} their binary counterpart. Our agents similarly pay a quadratic cost for stating an opinion other than the one they hold. We differ in that our agents care about the \emph{final} network consensus (rather than local disagreement), can report different opinions on different layers, and in that we compare the equilibrium consensus against the truthful one.

\emph{Games on multiplex and multilayer networks.} We adopt the multilayer network formalism of \cite{kivela2014multilayer,dedomenico2013mathematical}. Prior work on dynamical processes on these networks has studied diffusion and the spectrum of the merged Laplacian~\cite{gomez2013diffusion,soleribalta2013spectral}, co-evolving opinions across layers~\cite{diakonova2014absorbing}, and, more recently, how multiplexing of ties affects diffusion and the returns to seeding~\cite{chandrasekhar2026multiplexing}. Our starting point is the recent work of \cite{shiu2026coordination}, which proposes merged-layer coordination games and establishes the convergence and stability of the resulting opinion dynamics under truthful reporting. We take these merged dynamics as given and ask what happens when reports are strategic; the leverage decomposition of Lemma~\ref{lem:aggregation}, which allows per-layer reports to be aggregated by the merged dynamics without assuming that the layers share a stationary distribution, does not appear in that work.

\emph{Strategic misreporting and aggregation.} The cost of lying that we consider follows the strategic communication literature: \cite{kartik2007credulity,kartik2009strategic} study a sender who misreports to a receiver at a cost increasing in the size of the lie, and find that reports are inflated in equilibrium; \cite{hagenbach2010strategic,galeotti2013strategic} place cheap-talk senders on a network. Relatedly, \cite{glaeser2005strategic} explain extremism as a strategic response to who is listening, and \cite{ali2020image} study stated versus true positions when statements carry image concerns. In contrast to these works, which consider one (or a few) senders facing a receiver, we consider a population of agents whose statements are averaged by a network, with a platform-chosen weighting of layers; the inflation of reports we find (Proposition~\ref{prop:posturing}) is the population analog of the sender's inflation in \cite{kartik2009strategic}, while the cross-layer signature of Proposition~\ref{prop:signature} has no single-receiver counterpart. Our work also relates to the literature on manipulation-resistant aggregation of reported values, where strategyproofness pushes toward medians~\cite{moulin1980strategyproofness} and, for regression, toward specific estimators~\cite{dekel2010incentive,perote2004strategyproof,chen2018strategyproof}, and where \cite{kephart2016revelation} study mechanism design with costly misreporting. We do not seek a strategyproof rule; the aggregator in our model is a fixed, linear network consensus, and the platform's only instrument is its attention to each layer.

Finally, our work shares a common feature with the literature on strategic responses to algorithms, including strategic classification~\cite{hardt2016strategic,milli2019social,hu2019disparate}, performative prediction~\cite{perdomo2020performative}, and its alternative microfoundations~\cite{jagadeesan2021alternative}, in that agents alter what an algorithm observes at a cost; in our own prior work~\cite{ebrahimi2025doubleedged}, we showed that behavioral responses to a classifier can help or hurt the firm. In these works, the outcome each agent games is their own classification; here, the outcome is a single collective consensus, so that agents' manipulations interact through an aggregate. Our game is thus an aggregative game~\cite{jensen2010aggregative,cornes2007aggregative} with linear-quadratic payoffs, similar to the network games of \cite{ballester2006whoswho,bramoulle2014strategic,galeotti2020targeting}, in which equilibrium actions are proportional to a centrality; the tilted centrality of Theorem~\ref{thm:equilibrium} plays the role that Bonacich centrality plays there, with manipulability (rather than complementarity) as the source of the tilt.

\section{Model}\label{sec:model}

\subsection{Multiplex averaging and the leverage decomposition}

A set $\Vc=\{1,\dots,N\}$ of agents interacts on $L$ layers indexed by $\Lc=\{1,\dots,L\}$: different platforms, channels, or venues over the same population. Layer $l$ is a weighted directed graph with row-stochastic influence matrix $W^{(l)}\in\R^{N\times N}$, where $W^{(l)}_{ij}$ is the weight agent $i$ places on agent $j$ on that layer. A platform assigns an attention weight $\alpha_l>0$ to each layer, $\sum_l\alpha_l=1$, and works with the merged matrix
\begin{equation}\label{eq:merged}
C(\alpha)=\sum_{l\in\Lc}\alpha_l W^{(l)}.
\end{equation}
This is the merged-layer coordination game of \citet{shiu2026coordination}: an agent minimizing the attention-weighted sum of squared disagreements with their neighbors on each layer,
\[
\tfrac12\sum_l\alpha_l\sum_j W^{(l)}_{ij}(z_i-z_j)^2 ,
\]
best-responds by updating to $z_i\leftarrow (C(\alpha)z)_i$. We assume throughout that $C(\alpha)$ is primitive; it suffices that one layer is primitive, since $C(\alpha)^k\ge \alpha_l^k (W^{(l)})^k$ entrywise. Let $\pi=\pi(\alpha)$ be the unique stationary distribution of $C(\alpha)$, $\pi^\top C(\alpha)=\pi^\top$, with $\pi_i>0$ for all $i$. Under sincere reporting, iterating $z\leftarrow C(\alpha)z$ from an opinion vector $\theta$ converges to the consensus $\bar\theta=\pi^\top\theta$ \cite{degroot1974reaching,golub2010naive}.

Our departure is that an agent need not feed the same number into every layer. Agent $i$ posts a report $x_i^{(l)}\in\R$ on each layer $l$; write $x^{(l)}=(x_1^{(l)},\dots,x_N^{(l)})^\top$ and $x=(x^{(l)})_{l\in\Lc}$. The platform initializes the merged dynamics from these reports, so the first round of averaging reads $z_i(1)=\sum_l\alpha_l\sum_j W^{(l)}_{ij}x_j^{(l)}$ and subsequent rounds are $z(t+1)=C(\alpha)z(t)$. The consensus this produces has a simple form.

\begin{lemma}[Leverage decomposition]\label{lem:aggregation}
The merged dynamics initialized at the per-layer reports $x$ converge to
\begin{equation}\label{eq:y}
y(x)=\sum_{i\in\Vc}\sum_{l\in\Lc}\omega_{i,l}\,x_i^{(l)},\qquad \omega_{i,l}:=\alpha_l\,\big(\pi^\top W^{(l)}\big)_i ,
\end{equation}
where the leverages satisfy $\omega_{i,l}\ge0$ and $\sum_{l\in\Lc}\omega_{i,l}=\pi_i$ for every $i$. If all layers share the stationary distribution $\pi$ (that is, $\pi^\top W^{(l)}=\pi^\top$ for all $l$), then $\omega_{i,l}=\alpha_l\pi_i$.
\end{lemma}

Intuitively, the leverage $\omega_{i,l}$ is the probability that the (stationary) random walk on the merged network arrives at agent $i$ through an edge of layer $l$; that is, it is the share of the network's long-run attention that reaches $i$ on that layer. An agent's centrality in the merged network is the sum of their leverages across layers, and layer $l$'s contribution to it is larger the more attention the platform pays to that layer, and the more of the flow into $i$ that layer carries. We note that when all layers share a stationary distribution, leverage factorizes into attention times centrality; this is the case in which merging changes the weights on the layers but not \emph{who} is central, and we use it whenever we want to isolate the effects of attention from those of re-centralization. We also note that if reports are \emph{linked}, i.e., $x_i^{(l)}=x_i$ for all $l$, then \eqref{eq:y} reduces to $y=\pi^\top x$, the standard DeGroot consensus of the (single) reported profile.

\subsection{Strategic reporting}

\begin{assumption}[Private opinions and per-layer reports]
Each agent $i$ holds a fixed private opinion $\theta_i\in\R$ and chooses reports $x_i=(x_i^{(1)},\dots,x_i^{(L)})\in\R^L$. Reports are chosen once and simultaneously; the platform then runs the merged dynamics.
\end{assumption}

\begin{assumption}[Consensus anticipation]
Agents know the aggregation rule \eqref{eq:y}, or equivalently the platform's attention weights and the leverages $\omega_{i,l}$.
\end{assumption}

\begin{assumption}[Layer-specific misreporting costs]
Posting $x_i^{(l)}$ on layer $l$ costs agent $i$ $\tfrac{c_{i,l}}{2}(x_i^{(l)}-\theta_i)^2$ with $c_{i,l}>0$.
\end{assumption}

The cost captures whatever makes it unpleasant or risky to say something one does not believe: reputational exposure on a platform where one is identifiable, the effort of maintaining a persona, the chance of being contradicted by one's own record. It differs across layers because the same person is a named professional on one platform and an anonymous handle on another, and because platforms verify identities and moderate content to different degrees. Agent $i$ minimizes
\begin{equation}\label{eq:utility}
U_i(x_i,x_{-i})=\frac12\big(y(x)-\theta_i\big)^2+\sum_{l\in\Lc}\frac{c_{i,l}}{2}\big(x_i^{(l)}-\theta_i\big)^2 .
\end{equation}
The first term is the agent's stake in the outcome: they want the consensus the platform acts on to be close to their view. The second is the price of posturing. Together they define the game $\mathcal{G}=\langle\Vc,(\R^L)_{i\in\Vc},(U_i)_{i\in\Vc}\rangle$. Since $y$ depends on $x$ only through a linear aggregate, $\mathcal{G}$ is an aggregative game with quadratic payoffs; we use this structure directly.

We measure the harm of manipulation by the \emph{distortion}
\begin{equation}\label{eq:pom}
\PoM(\alpha)=\big|y^*-\bar\theta\big|,
\end{equation}
the distance between the equilibrium consensus $y^*$ and the sincere consensus $\bar\theta=\pi(\alpha)^\top\theta$ that the same platform would have computed had everyone reported truthfully. The benchmark is the platform's own sincere output, not the unweighted mean of opinions, so distortion isolates the effect of strategic reporting from the platform's choice of whom to listen to.

\section{Equilibrium and the distortion identity}\label{sec:equilibrium}

Two summary statistics drive the analysis. The \emph{manipulability index} of agent $i$ and the total manipulability of the network are
\begin{equation}\label{eq:K}
K_i=\sum_{l\in\Lc}\frac{\omega_{i,l}^2}{c_{i,l}},\qquad K=\sum_{i\in\Vc}K_i .
\end{equation}
Intuitively, $K_i$ measures how much agent $i$ can move the consensus per unit of misreporting cost: their squared leverage on each layer, discounted by their cost on that layer, and summed over the layers on which they can act.

\begin{theorem}[Equilibrium]\label{thm:equilibrium}
The game $\mathcal{G}$ has a unique Nash equilibrium. The equilibrium consensus is the sincere consensus under the tilted centrality $\tilde\pi_i=(\pi_i+K_i)/(1+K)$,
\begin{equation}\label{eq:ystar}
y^*=\sum_{i\in\Vc}\tilde\pi_i\,\theta_i=\frac{\bar\theta+\sum_i K_i\theta_i}{1+K},
\end{equation}
and agent $i$'s report on layer $l$ is
\begin{equation}\label{eq:xstar}
x_i^{(l)*}=y^*+\Big(1+\frac{\omega_{i,l}}{c_{i,l}}\Big)\big(\theta_i-y^*\big).
\end{equation}
\end{theorem}

\emph{Intuitive interpretation.} Theorem~\ref{thm:equilibrium} states that strategic manipulation re-weights the network: while the platform believes it is listening to agents in proportion to their centralities $\pi$, at equilibrium it is in fact listening to them in proportion to $\pi+K$ (normalized). In other words, an agent who is central and can lie cheaply on a layer that the platform pays attention to counts for more than their centrality, and the consensus moves toward them. Equation \eqref{eq:xstar} further shows \emph{how} agents achieve this: each agent takes their distance from the consensus, $\theta_i-y^*$, and stretches it by a factor $1+\omega_{i,l}/c_{i,l}$ on layer $l$. That is, no agent reports an opinion on the other side of the consensus from their own, and no agent reports an opinion closer to the consensus than the one they hold. We also note that the stretch is largest on the layer on which the agent's leverage-to-cost ratio is largest, so that the same agent's reports differ across layers; we return to this in Section~\ref{sec:observables}.

We next compare the equilibrium consensus $y^*$ with its truthful counterpart $\bar\theta$. For weights $\pi$ write $\Cov_\pi(u,v)=\sum_i\pi_i(u_i-\pi^\top u)(v_i-\pi^\top v)$ and $\sd_\pi(u)=\Cov_\pi(u,u)^{1/2}$, and let $g_i=K_i/\pi_i$ be manipulability per unit of centrality.

\begin{theorem}[Distortion identity]\label{thm:distortion}
In equilibrium,
\begin{equation}\label{eq:distortion}
y^*-\bar\theta=\frac{\sum_i K_i(\theta_i-\bar\theta)}{1+K}=\frac{\Cov_\pi(g,\theta)}{1+K}.
\end{equation}
Consequently: (i) $y^*=\bar\theta$ if and only if $g$ and $\theta$ are uncorrelated under $\pi$; (ii) $\PoM(\alpha)\le \sd_\pi(g)\,\sd_\pi(\theta)/(1+K)$, with equality if and only if $\theta-\bar\theta\mathbf 1$ is proportional to $g-K\mathbf 1$; (iii) over all opinion profiles with $\sd_\pi(\theta)\le\sigma$, the largest distortion is $\sigma\,\sd_\pi(g)/(1+K)$, attained when opinions are aligned with manipulability.
\end{theorem}

\emph{Intuitive interpretation.} Theorem~\ref{thm:distortion} states that cheap lying is not, by itself, harmful: if the agents who can move the consensus hold opinions that are typical (under $\pi$), their manipulations cancel out, and the platform recovers the truthful consensus. Distortion arises only when manipulability and opinions \emph{covary}, i.e., when the agents with leverage on cheap layers are also the ones whose opinions are far from the mean. Part (ii) separates the two ingredients of this covariance: the dispersion of manipulability, $\sd_\pi(g)$, is a property of the network and the cost structure, and is (partially) under the platform's control through $\alpha$, whereas the dispersion of opinions is not. Part (iii) identifies the worst case that a platform which knows the leverages and costs, but not the opinions, should plan for; we use it in Section~\ref{sec:design}.

\section{What the platform observes}\label{sec:observables}

In practice, a platform rarely observes agents' opinions; it observes their reports. We next present two consequences of \eqref{eq:xstar} regarding the gap between the two.

\begin{proposition}[Posturing]\label{prop:posturing}
Suppose opinions are not all equal. Then $\min_i\theta_i<y^*<\max_i\theta_i$, and for every agent $i$ and layer $l$, $x_i^{(l)*}-y^*$ has the sign of $\theta_i-y^*$ and $|x_i^{(l)*}-y^*|\ge|\theta_i-y^*|$, strictly whenever $\theta_i\ne y^*$ and $\omega_{i,l}>0$. In particular, on every layer $l$ on which the agents holding the most extreme opinions have positive leverage,
\begin{equation}
\max_i x_i^{(l)*}-\min_i x_i^{(l)*}\;>\;\max_i\theta_i-\min_i\theta_i .
\end{equation}
\end{proposition}

\emph{Intuitive interpretation.} Proposition~\ref{prop:posturing} states that the distribution of reports observed on each layer is a (pointwise) stretch of the distribution of opinions, away from the consensus. Consequently, the range of reports, and in fact any measure of dispersion around the consensus, is inflated on every layer, and it is inflated the most for agents with high leverage and low costs. A platform that estimates polarization from reports will therefore overestimate it, and will do so most for exactly the agents who distort the consensus. We note that the proposition also identifies the \emph{direction} of this bias: the most extreme reports come from the agents with the most extreme opinions, so that the observed polarization is exaggerated by the wings rather than fabricated by the moderates.

\begin{proposition}[Cross-layer signature]\label{prop:signature}
For every agent $i$ with $\theta_i\ne y^*$ and every pair of layers $l,m$,
\begin{align}
\frac{x_i^{(l)*}-y^*}{x_i^{(m)*}-y^*}&=\frac{1+\omega_{i,l}/c_{i,l}}{1+\omega_{i,m}/c_{i,m}},\label{eq:ratio}\\
x_i^{(l)*}-x_i^{(m)*}&=(\theta_i-y^*)\Big(\frac{\omega_{i,l}}{c_{i,l}}-\frac{\omega_{i,m}}{c_{i,m}}\Big).\label{eq:diff}
\end{align}
An agent's reports are ordered across layers by $\omega_{i,l}/c_{i,l}$, with the most extreme report on the layer where leverage per unit cost is largest, and the ratio of an agent's distances from the consensus on two layers does not depend on $\theta_i$.
\end{proposition}

\emph{Intuitive interpretation.} Equation~\eqref{eq:ratio} can be viewed as an identification result: a platform that observes an agent's reports on two layers, together with its own consensus, can recover the agent's relative leverage-to-cost ratio on those layers, without knowing the agent's opinion. Since the platform knows the leverages (it chose $\alpha$ and observes the network), it can in turn recover the agent's relative misreporting costs. Equation~\eqref{eq:diff}, on the other hand, states that the inconsistency of an agent's reports across layers is proportional to their distance from the consensus. In other words, cross-layer inconsistency is not noise; it flags the agents who are both far from the consensus and able to act on it, which are exactly the agents that Theorem~\ref{thm:distortion} identifies as the source of distortion.

\section{Does merging layers help?}\label{sec:multiplexing}

A platform that merges more layers observes more of the population and, under truthful reporting, can benefit from the improved connectivity of the merged network~\cite{shiu2026coordination}. Under strategic reporting, however, one may expect that merging layers also gives agents more room to manipulate the consensus. In this section, we show that this is not necessarily the case, and that whether merging helps or hurts is determined by the square in \eqref{eq:K}.

\subsection{Dilution}

We first isolate the effect of the attention weights from that of centrality, by assuming that the layers share a stationary distribution, so that $\omega_{i,l}=\alpha_l\pi_i$ (Lemma~\ref{lem:aggregation}). In this case, $K_i(\alpha)=\pi_i^2\sum_l\alpha_l^2/c_{i,l}$, so that the attention weights enter only through their squares.

\begin{theorem}[Dilution]\label{thm:dilution}
Assume all layers share the stationary distribution $\pi$.
\begin{enumerate}
\item[(a)] If costs are layer-uniform, $c_{i,l}=c_i$, then $K_i(\alpha)=s(\alpha)\,\pi_i^2/c_i$ with $s(\alpha)=\sum_l\alpha_l^2\in[1/L,1]$, and
\begin{equation}
y^*(\alpha)-\bar\theta=\frac{s(\alpha)\,D}{1+s(\alpha)\,K^{(1)}},\qquad D=\sum_i\frac{\pi_i^2}{c_i}(\theta_i-\bar\theta),\ K^{(1)}=\sum_i\frac{\pi_i^2}{c_i}.
\end{equation}
Distortion is strictly increasing in $s(\alpha)$: it is largest when the platform uses a single layer and smallest, by a factor $L$ in the numerator, under uniform attention.
\item[(b)] For arbitrary costs, let the platform add a new layer $L+1$ at weight $\varepsilon$ and rescale the existing weights by $1-\varepsilon$. If $\PoM(0)>0$, then
\begin{equation}
\frac{d}{d\varepsilon}\log\PoM(\varepsilon)\Big|_{\varepsilon=0}=-\frac{2}{1+K(0)}<0 ,
\end{equation}
so adding any layer at a small enough weight strictly reduces distortion. At full weight, $\PoM(1)>\PoM(0)$ if and only if $|V_{L+1}|/(1+S_{L+1})>\PoM(0)$, where
\[
V_{L+1}=\sum_i\frac{\pi_i^2(\theta_i-\bar\theta)}{c_{i,L+1}},\qquad S_{L+1}=\sum_i\frac{\pi_i^2}{c_{i,L+1}} .
\]
\end{enumerate}
\end{theorem}

\emph{Intuitive interpretation.} Part (a) of Theorem~\ref{thm:dilution} runs against the intuition that more layers mean more opportunity to manipulate. Intuitively, with uniform costs, an agent who misreports on $L$ layers pays $L$ separate (convex) costs, while the influence they buy on each layer is scaled by $\alpha_l$; as leverage enters $K_i$ as a square, splitting the platform's attention across layers reduces every agent's manipulability by the Herfindahl index $s(\alpha)$ of the attention profile. Part (b) makes the same point without assuming uniform costs: adding a new layer at weight $\varepsilon$ scales every existing leverage by $1-\varepsilon$, and hence every existing manipulability by $(1-\varepsilon)^2$ (a first-order reduction), whereas the new layer only contributes manipulability of order $\varepsilon^2$. Perhaps unexpectedly, therefore, however cheap the new layer is for extreme agents, adding a little of it reduces distortion (whenever there was distortion to begin with). Merging hurts only if the platform gives the new layer enough weight that the layer's own vulnerability, $|V_{L+1}|/(1+S_{L+1})$, exceeds the distortion the platform started with; we illustrate the threshold weight at which this happens in Section~\ref{sec:numerics}.

\subsection{Cost heterogeneity and identity linking}

The first way in which merging can hurt is that reporting separately on each layer lets an agent concentrate their manipulation on the layers where lying is cheap. To assess this, we compare against \emph{linked} reporting, in which the platform (or the identity infrastructure across platforms) forces $x_i^{(l)}=x_i$ for all $l$. A linked agent still pays the cost of every layer, $\sum_l\tfrac{c_{i,l}}{2}(x_i-\theta_i)^2$, and their leverage is $\pi_i$, so that their manipulability is $K_i^{\mathrm{lk}}=\pi_i^2/\sum_l c_{i,l}$. The following result requires no assumption on the layers' stationary distributions.

\begin{theorem}[Linking]\label{thm:linking}
For every agent $i$ and every attention profile,
\begin{equation}
K_i=\sum_l\frac{\omega_{i,l}^2}{c_{i,l}}\;\ge\;\frac{\big(\sum_l\omega_{i,l}\big)^2}{\sum_l c_{i,l}}=\frac{\pi_i^2}{\sum_l c_{i,l}}=K_i^{\mathrm{lk}},
\end{equation}
with equality if and only if $\omega_{i,l}/c_{i,l}$ is constant across layers. Unlinked reporting weakly raises every agent's manipulability and the total $K$, and the gap for agent $i$ is
$\sum_{l<m}\big(\omega_{i,l}c_{i,m}-\omega_{i,m}c_{i,l}\big)^2/\big(c_{i,l}c_{i,m}\sum_{l'}c_{i,l'}\big)$.
\end{theorem}

\emph{Intuitive interpretation.} The gap in Theorem~\ref{thm:linking} vanishes exactly when the agent has no reason to say different things on different layers, i.e., when the layer that gives them the most leverage is also (proportionally) the layer on which lying is most expensive. In every other case, the freedom to tailor reports to each layer is worth something to the manipulator. This result points to an intervention that our model can evaluate directly: cross-platform identity verification, or any mechanism that makes an agent's reports on one layer attributable to their reports on another, closes the Cauchy-Schwarz gap and returns every agent to their linked manipulability. We note, however, that as distortion is a (signed) covariance rather than a monotone function of the individual $K_i$'s (Theorem~\ref{thm:distortion}), linking reduces every agent's leverage over the consensus, but need not reduce the distortion in every instance; it does so whenever it scales all $K_i$'s by a common factor, as in Theorem~\ref{thm:dilution}(a), and we show an instance in which it does not in Section~\ref{sec:numerics}.

\subsection{Re-centralization}

The second way in which merging can hurt is unrelated to costs. When the layers differ in their centrality structure, $\pi(\alpha)$ changes with $\alpha$, and a new layer can shift leverage toward agents whose opinions are far from the mean. As the following example shows, this effect can overwhelm dilution even at small weights.

\begin{example}[Re-centralization]\label{ex:recentralization}
Four agents hold opinions $\theta=(0,\,0.1,\,-0.1,\,1)$ with unit costs on both layers. On layer $A$, agents $2$, $3$, $4$ each place weight $0.8$ on agent $1$ and the remainder on agent $4$ (agent $4$ splits its remainder between $2$ and $3$), and agent $1$ listens to $2$, $3$, $4$ with weights $(0.4,0.4,0.2)$: agent $1$, a moderate, is the hub. On layer $B$, agents $1$, $2$, $3$ place weight $0.8$ on agent $4$ and $0.2$ on agent $1$, and agent $4$ listens uniformly to the others: the extremist is the hub. Adding layer $B$ at weight $\varepsilon$ raises the extremist's centrality from $\pi_4=0.17$ at $\varepsilon=0$ to $0.44$ at $\varepsilon=1$. Distortion rises from $0.017$ at $\varepsilon=0$ to $0.022$ at $\varepsilon=0.11$ (Figure~\ref{fig:recentralization}): the dilution term of Theorem~\ref{thm:dilution}(b), which is present here too, is dominated by the shift of leverage toward agent $4$. Section~\ref{sec:general} quantifies the two effects and shows that the same instance has an interior attention weight at which distortion is exactly zero.
\end{example}

Taken together, merging layers protects the consensus by default (Theorem~\ref{thm:dilution}), and can harm it through two channels that a platform can assess before merging: differences in misreporting costs across layers, which identity linking removes (Theorem~\ref{thm:linking}), and shifts of centrality toward agents with extreme opinions, which the leverage decomposition exposes (Example~\ref{ex:recentralization}).

\section{Choosing the attention weights}\label{sec:design}

We now consider a platform that chooses its attention weights $\alpha$ so as to minimize distortion. We first solve the case in which $\alpha$ does not change who is central, for which we obtain a closed-form solution, and then turn to the general case, in which re-centralization enters and a closed form is no longer available; for the latter, we provide an exact gradient, a test for whether adding a layer helps at the margin, and an exact algorithm for the two-layer case.

\subsection{Layers with a common stationary distribution}

Define, for each layer, the \emph{vulnerability score} and the \emph{layer manipulability}
\begin{equation}
V_l=\sum_i\frac{\pi_i^2}{c_{i,l}}(\theta_i-\bar\theta),\qquad S_l=\sum_i\frac{\pi_i^2}{c_{i,l}} ,
\end{equation}
so that $\sum_iK_i(\theta_i-\bar\theta)=\sum_l\alpha_l^2V_l$ and $K=\sum_l\alpha_l^2S_l$. Intuitively, $V_l$ is positive (resp. negative) when the agents who can lie cheaply on layer $l$ lean above (resp. below) the truthful consensus, and its magnitude is the centrality-weighted covariance of opinions and cheapness on that layer. The platform's problem is
\begin{equation}\label{eq:design}
\min_{\alpha\in\Delta}\ \PoM(\alpha)=\frac{\big|\sum_l\alpha_l^2V_l\big|}{1+\sum_l\alpha_l^2S_l}.
\end{equation}

\begin{theorem}[Robust attention]\label{thm:design}
Assume all layers share the stationary distribution $\pi$.
\begin{enumerate}
\item[(a)] If some $V_l=0$, or the $V_l$ are not all of the same sign, then zero distortion is attainable, and the set of zero-distortion weightings is $\{\alpha\in\Delta:\sum_l\alpha_l^2V_l=0\}$, a nonempty subset of the simplex.
\item[(b)] If all $V_l>0$ (the case $V_l<0$ is symmetric), the minimizer of \eqref{eq:design} is unique and equals
\begin{equation}\label{eq:alphastar}
\alpha_l^*=\frac{(V_l-f^*S_l)^{-1}}{\sum_m(V_m-f^*S_m)^{-1}},
\end{equation}
where $f^*=\min_\alpha\PoM(\alpha)$ is the unique root in $(0,\min_l V_l/S_l)$ of
\begin{equation}\label{eq:root}
f\sum_{l}\frac{1}{V_l-fS_l}=1 .
\end{equation}
In the small-manipulation limit $S_l\to0$, $\alpha_l^*\to V_l^{-1}/\sum_mV_m^{-1}$.
\end{enumerate}
\end{theorem}

\emph{Intuitive interpretation.} Part (a) of Theorem~\ref{thm:design} is a cancellation result: if one layer empowers agents who lean one way and another layer empowers agents who lean the other, the platform can set its attention weights such that the two vulnerabilities offset, and recover the truthful consensus. We note that this requires neither silencing any agent, nor knowledge of individual opinions; it only requires the signs and magnitudes of $L$ layer-level scores. Part (b) covers the case in which all layers lean the same way, so that no cancellation is possible. Here, the optimal rule weights each layer inversely to its vulnerability, corrected by a term proportional to the layer's manipulability, where the proportionality constant is the minimal distortion itself. Intuitively, the correction works in a definite direction: a layer with high $S_l$ receives \emph{more} weight than the naive inverse-vulnerability rule would assign to it, because total manipulability enters the denominator of distortion, and a heavily manipulated layer pulls the consensus toward a manipulation-weighted average that is less sensitive to any one agent. The naive rule is recovered when manipulation is small. We also note that a platform which instead weights layers by, e.g., their traffic or connectivity, ignores $V_l$ altogether and may land anywhere on the distortion curve.

\subsection{General layers}\label{sec:general}

We now drop the assumption of a common stationary distribution. In this case, $\pi=\pi(\alpha)$, the leverages $\omega_{i,l}(\alpha)=\alpha_l(\pi(\alpha)^\top W^{(l)})_i$ and the truthful benchmark $\bar\theta(\alpha)=\pi(\alpha)^\top\theta$ all change with $\alpha$, and the numerator of distortion is no longer quadratic in $\alpha$. The main technical challenge is that there is no closed-form expression for how the stationary distribution of the merged network $C(\alpha)$ depends on $\alpha$. We however show that three properties survive. First, distortion has an exact gradient, as the stationary distribution of a primitive chain is differentiable in the chain, with a derivative given by a simple formula through the fundamental matrix $Z(\alpha)=(I-C(\alpha)+\mathbf 1\pi^\top)^{-1}$. Second, this derivative decomposes into the dilution term of Theorem~\ref{thm:dilution} and a re-centralization term, which yields a test for whether adding a layer helps at the margin. Third, $\pi(\alpha)$ is a ratio of polynomials in $\alpha$ of degree at most $N-1$, so that the design problem is a polynomial fractional program: whether zero distortion is attainable is decided by the sign of a single polynomial, and for two layers, the global optimum can be found exactly by univariate root-finding.

Write $N(\alpha)=\sum_iK_i(\theta_i-\bar\theta)$ and $Q(\alpha)=1+K$, so $\PoM=|N|/Q$, and for a direction $d\in\R^L$ with $\sum_ld_l=0$ write $\dot u$ for the directional derivative of a quantity $u$ along $d$.

\begin{theorem}[Gradient of distortion]\label{thm:gradient}
Let $\alpha$ be a point of the simplex at which $C(\alpha)$ is primitive and $N(\alpha)\ne0$, and let $d$ be a direction with $\sum_ld_l=0$ that points into the simplex (the derivative is one-sided at a boundary point). Then, with $\dot C=\sum_ld_lW^{(l)}$,
\begin{align}
\dot\pi^\top&=\pi^\top\dot C\,Z(\alpha),\qquad
\dot\omega_{i,l}=d_l(\pi^\top W^{(l)})_i+\alpha_l(\dot\pi^\top W^{(l)})_i,\label{eq:pidot}\\
\dot K_i&=\sum_l\frac{2\omega_{i,l}\dot\omega_{i,l}}{c_{i,l}},\quad
\dot N=\sum_i\dot K_i(\theta_i-\bar\theta)-K\dot\pi^\top\theta,\quad \dot Q=\sum_i\dot K_i,\label{eq:Ndot}
\end{align}
and $\frac{d}{d\varepsilon}\log\PoM(\alpha+\varepsilon d)\big|_{0}=\dot N/N-\dot Q/Q$. In particular, adding a new layer $L+1$ at weight $\varepsilon$ and rescaling the others by $1-\varepsilon$ gives
\begin{align}
\frac{d}{d\varepsilon}\log\PoM(\varepsilon)\Big|_{\varepsilon=0}&=-\frac{2}{1+K}+\mathcal R,\label{eq:marginal}\\
\mathcal R&=\frac{\sum_i\rho_i(\theta_i-\bar\theta)-K\,\dot\pi^\top\theta}{N}-\frac{\sum_i\rho_i}{1+K},\notag
\end{align}
where
\[
\dot\pi^\top=\pi^\top\big(W^{(L+1)}-C(\alpha)\big)Z(\alpha),\qquad
\rho_i=\sum_l\frac{2\omega_{i,l}\alpha_l(\dot\pi^\top W^{(l)})_i}{c_{i,l}} .
\]
The new layer raises distortion at the margin if and only if $\mathcal R>2/(1+K)$, and $\mathcal R=0$ whenever $\dot\pi=0$, in which case \eqref{eq:marginal} reduces to Theorem~\ref{thm:dilution}(b).
\end{theorem}

\emph{Intuitive interpretation.} The re-centralization term $\mathcal R$ consists of three parts, each of which can be computed from the network, the costs, and the opinions: the change in agents' manipulability as leverage moves between them (the $\rho_i$ terms in the numerator), the movement of the truthful benchmark itself ($K\dot\pi^\top\theta$), and the change in total manipulability (in the denominator). Theorem~\ref{thm:gradient} thus turns the qualitative story of Example~\ref{ex:recentralization} into a number: in that example, the dilution term is $-1.54$ while $\mathcal R=7.21$, i.e., the new layer shifts leverage toward the extreme agent much faster than it dilutes it, and distortion rises. In practice, a platform deciding whether to merge a new channel can evaluate \eqref{eq:marginal}, at the cost of a single linear solve for $Z$, before merging.

\begin{theorem}[Algebraic structure and exact solution]\label{thm:algebraic}
Assume every layer is primitive.
\begin{enumerate}
\item[(a)] There are polynomials $a_1,\dots,a_N$ in $\alpha$ of degree at most $N-1$, with $s=\sum_ja_j>0$ on the simplex, such that $\pi_i(\alpha)=a_i(\alpha)/s(\alpha)$. Consequently $\PoM(\alpha)=|P(\alpha)|/R(\alpha)$ for polynomials $P,R$ of degree at most $3N-1$ with $R>0$ on the simplex.
\item[(b)] Zero distortion is attainable if and only if $P$ has a zero on the simplex. It is attainable whenever the single-layer vulnerabilities
$V^{(l)}=\sum_i\big(\pi^{(l)}_i\big)^2\big(\theta_i-\pi^{(l)\top}\theta\big)/c_{i,l}$, computed under each layer's own stationary distribution $\pi^{(l)}$, are not all of the same sign, or some $V^{(l)}=0$.
\item[(c)] For $L=2$, writing $\alpha=(1-\varepsilon,\varepsilon)$, a global minimizer of $\PoM$ over $[0,1]$ lies in the finite set consisting of $0$, $1$, the real zeros of $P$ in $(0,1)$, and the real zeros of $P'R-PR'$ in $(0,1)$, a polynomial of degree at most $6N-3$. The global minimum is therefore computable exactly by univariate root isolation.
\end{enumerate}
\end{theorem}

\begin{figure}[t]
\centering
\includegraphics[width=0.82\linewidth]{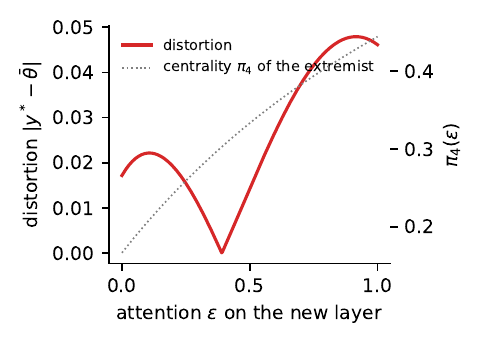}
\caption{Re-centralization (Example~\ref{ex:recentralization}). The new layer makes the extremist central; distortion rises at small weights despite dilution (Theorem~\ref{thm:gradient}: $\mathcal R=7.21>2/(1+K)=1.54$), vanishes at $\varepsilon^*=0.391$ where the numerator polynomial changes sign (Theorem~\ref{thm:algebraic}), and rises again as the extremist's centrality dominates.}
\label{fig:recentralization}
\end{figure}

\emph{Intuitive interpretation.} Part (a) of Theorem~\ref{thm:algebraic} shows where re-centralization is paid for: with a common stationary distribution, the numerator of distortion is a quadratic form in $\alpha$, which Theorem~\ref{thm:design} minimizes in closed form; in general, it is a polynomial whose degree grows with the number of agents, and the closed form \eqref{eq:alphastar} does not carry over. What remains true is that the platform's problem is a low-dimensional polynomial optimization over a simplex, and that for the question of whether, and how much, to merge one new channel into an existing aggregate, the answer is exact. Part (b) generalizes the cancellation result of Theorem~\ref{thm:design}(a), with the one change that re-centralization forces: each layer's vulnerability is now measured under that layer's \emph{own} centrality, as at a vertex of the simplex that is who is central. We note that this condition is only sufficient in general, as $P$ may change sign inside the simplex without changing sign at its vertices. In Example~\ref{ex:recentralization}, the vertex signs differ, and part (c) locates the zero of $P$ at $\varepsilon^*=0.391$: a platform that merges the extremist-centered layer at about $39\%$ attention recovers the truthful consensus, even though merging it at $5\%$ makes things worse, and merging it fully is worse still (Figure~\ref{fig:recentralization}).

\section{Numerical illustration}\label{sec:numerics}

\begin{figure}[t]
\centering
\includegraphics[width=0.82\linewidth]{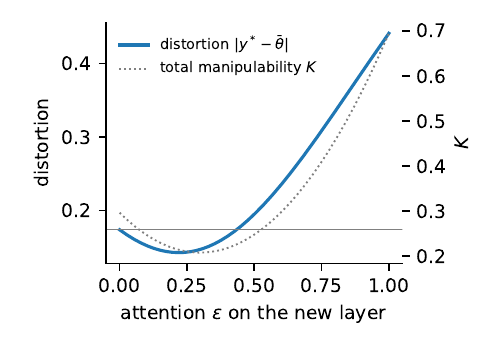}
\caption{Adding a layer that is cheap for the most extreme agent (Example~\ref{ex:dilution}). Distortion falls at small attention weights and rises past its single-layer value only once the new layer receives about $44\%$ of attention; total manipulability $K$ falls and then rises with it.}
\label{fig:dilution}
\end{figure}

\begin{example}[Dilution then harm]\label{ex:dilution}
Six agents hold opinions $\theta=(-1,-0.6,-0.2,0.2,0.6,1.5)$; agent $6$ is far from the mean. Layer $1$ is a symmetric ring in which each agent places weight $0.3$ on each immediate neighbor and $0.2$ on each neighbor at distance two; layer $2$ is a symmetric ring with weights $0.2$ on immediate neighbors and $0.3$ on neighbors at distance three. Both are doubly stochastic, so $\pi$ is uniform for every $\alpha$ and Theorems~\ref{thm:dilution} and \ref{thm:design} apply. Costs on layer $1$ are $(1,2,1,1,2,0.15)$: agent $6$ is already cheap to move. Layer $2$ has unit costs except $c_{6,2}=0.05$: it is even cheaper for the extremist. Using layer $1$ alone gives distortion $0.174$. Figure~\ref{fig:dilution} plots distortion as attention $\varepsilon$ shifts to layer $2$. It falls to $0.143$ at $\varepsilon=0.22$, returns to its single-layer value at $\varepsilon\approx0.44$, and reaches $0.441$ at $\varepsilon=1$; the initial slope is $-2\PoM(0)/(1+K(0))$ as in Theorem~\ref{thm:dilution}(b). At $\varepsilon=0.3$ the sincere consensus is $0.083$ and the equilibrium consensus is $0.231$; the range of opinions is $2.5$ while the range of reports is $3.63$ on layer $1$ and $3.83$ on layer $2$, and agent $6$ reports $2.49$ on layer $1$ and $2.77$ on layer $2$, consistent with Propositions~\ref{prop:posturing} and \ref{prop:signature}. Linking profiles in this instance yields distortion $0.148$ at every $\varepsilon$: below the single-layer value but above the unlinked minimum, which illustrates the caveat after Theorem~\ref{thm:linking}. Both layers have $V_l>0$ here ($V=(0.225,0.748)$, $S=(0.296,0.694)$), so Theorem~\ref{thm:design}(b) applies: the root of \eqref{eq:root} is $f^*=0.143$, the optimal attention on layer $2$ is $\alpha_2^*=0.220$, matching the minimizer in the figure, and the naive inverse-vulnerability rule would give $0.232$.
\end{example}


\section{Conclusion}\label{sec:conclusion}

We have proposed a model of strategic opinion manipulation on multiplex networks, in which a platform merges several network layers into one consensus, and agents, who know that they are being aggregated, choose the opinions they report on each layer strategically. This model enabled us to explore how strategic reporting distorts the consensus, what the platform observes, and whether merging layers supports or undermines the robustness of the consensus to manipulation. At a technical level, the game reduces to an aggregative game with a linear-quadratic structure, which yields a unique equilibrium in closed form: the platform computes the truthful consensus under a centrality tilted toward agents' manipulability, and the distortion is a covariance of manipulability and opinions. Answering the design questions required us to understand how agents' leverage, and the stationary distribution of the merged network, depend on the platform's attention weights; for the former, we showed that leverage enters manipulability as a square, and for the latter, we used the derivative of the stationary distribution and the polynomial structure of the merged chain, as no closed-form expression is available in general.

Our findings shed light on the reasons why aggregating opinions across networks can be fragile, and can guide potential interventions. For instance, we noted that merging layers dilutes manipulation by default, and that it becomes harmful only through differences in misreporting costs across layers, or through shifts of centrality toward agents with extreme opinions; this suggests interventions in which a platform links agents' identities across channels (removing the former effect), or, before merging a new channel, assesses the re-centralization it induces against the dilution it brings (Theorem~\ref{thm:gradient}). We also noted that the polarization a platform measures from reports overestimates the true polarization, most for the agents who distort the consensus, and that the inconsistency of an agent's reports across layers identifies their leverage; these observations may be useful in estimating misreporting costs from multi-platform data.

We conclude with some potential directions of future work. First, our model considers quadratic payoffs and unconstrained reports, which is what makes the equilibrium linear; bounded report spaces (e.g., discrete ratings) would introduce corner solutions that remain to be analyzed. Second, agents in our model know the platform's aggregation rule; a version in which agents hold (possibly biased) beliefs about the attention weights would connect our model to behavioral responses to algorithms~\cite{ebrahimi2025doubleedged}. Third, reports are chosen once; a repeated version in which agents update their reports as the consensus evolves would connect to opinion games with stubborn agents~\cite{bindel2015forming}. Finally, our closed-form design rule requires the layers to share a stationary distribution, and for three or more general layers, the platform's problem remains a polynomial fractional program; identifying conditions under which a closed form is recovered remains an open direction.

\bibliographystyle{plainnat}
\bibliography{references}

\appendix

\section{Proofs}\label{app:proofs}

\subsection{Proof of Lemma~\ref{lem:aggregation}}
After the first round, $z(1)=\sum_l\alpha_lW^{(l)}x^{(l)}$ and $z(t)=C(\alpha)^{t-1}z(1)$ for $t\ge1$. Since $C(\alpha)$ is primitive and row-stochastic, $C(\alpha)^{t}\to\mathbf 1\pi^\top$, so $z(t)\to\mathbf 1\,\pi^\top z(1)=\mathbf 1\sum_l\alpha_l\,\pi^\top W^{(l)}x^{(l)}$, which is \eqref{eq:y} with $\omega_{i,l}=\alpha_l(\pi^\top W^{(l)})_i$. Nonnegativity is immediate. Summing over $l$, $\sum_l\omega_{i,l}=(\pi^\top\sum_l\alpha_lW^{(l)})_i=(\pi^\top C(\alpha))_i=\pi_i$. If $\pi^\top W^{(l)}=\pi^\top$ for all $l$ then $\omega_{i,l}=\alpha_l\pi_i$. \qed

\subsection{Proof of Theorem~\ref{thm:equilibrium}}
$U_i$ is a strictly convex quadratic in $x_i\in\R^L$ (its Hessian is $\omega_i\omega_i^\top+\mathrm{diag}(c_{i,\cdot})\succ0$), so agent $i$'s best response to any $x_{-i}$ is the unique solution of the first-order conditions
\begin{equation}\label{eq:foc}
\omega_{i,l}\big(y(x)-\theta_i\big)+c_{i,l}\big(x_i^{(l)}-\theta_i\big)=0,\qquad l\in\Lc .
\end{equation}
A profile is a Nash equilibrium if and only if \eqref{eq:foc} holds for all $i,l$. Fix such a profile and write $y=y(x)$. Then $x_i^{(l)}=\theta_i-(\omega_{i,l}/c_{i,l})(y-\theta_i)$, and substituting into \eqref{eq:y},
\[
y=\sum_{i,l}\omega_{i,l}\theta_i-\sum_{i,l}\frac{\omega_{i,l}^2}{c_{i,l}}(y-\theta_i)=\bar\theta-Ky+\sum_iK_i\theta_i ,
\]
using $\sum_l\omega_{i,l}=\pi_i$. Since $1+K>0$ this has the unique solution \eqref{eq:ystar}, and the reports are then uniquely determined by \eqref{eq:foc}, which rearranges to \eqref{eq:xstar}. Conversely the profile so defined satisfies \eqref{eq:foc} with $y(x)=y^*$, so it is an equilibrium. The tilted weights $\tilde\pi_i=(\pi_i+K_i)/(1+K)$ are positive and sum to one. \qed

\subsection{Proof of Theorem~\ref{thm:distortion}}
From \eqref{eq:ystar}, $(1+K)(y^*-\bar\theta)=\sum_iK_i\theta_i-K\bar\theta=\sum_iK_i(\theta_i-\bar\theta)$. Writing $K_i=\pi_ig_i$ and using $\sum_i\pi_i(\theta_i-\bar\theta)=0$,
\[
\sum_i\pi_ig_i(\theta_i-\bar\theta)=\sum_i\pi_i(g_i-\pi^\top g)(\theta_i-\bar\theta)=\Cov_\pi(g,\theta),
\]
which is \eqref{eq:distortion}. Part (i) is immediate. Part (ii) is the Cauchy-Schwarz inequality for the inner product $\langle u,v\rangle=\sum_i\pi_iu_iv_i$ applied to the centered vectors, with equality if and only if they are proportional. Part (iii): $|\Cov_\pi(g,\theta)|\le\sd_\pi(g)\sigma$ over the stated set, with equality at $\theta=\bar\theta\mathbf 1+\sigma(g-K\mathbf 1)/\sd_\pi(g)$. \qed

\subsection{Proof of Proposition~\ref{prop:posturing}}
By Theorem~\ref{thm:equilibrium}, $y^*$ is a convex combination of the $\theta_i$ with strictly positive weights $\tilde\pi_i$ (positive because $\pi_i>0$ and $K_i\ge0$), so if opinions are not all equal, $\min_i\theta_i<y^*<\max_i\theta_i$. By \eqref{eq:xstar}, $x_i^{(l)*}-y^*=(1+\omega_{i,l}/c_{i,l})(\theta_i-y^*)$ with $1+\omega_{i,l}/c_{i,l}\ge1$, strictly if $\omega_{i,l}>0$, which gives the sign and magnitude claims. Let $H\in\arg\max_i\theta_i$ and $M\in\arg\min_i\theta_i$ have positive leverage on layer $l$. Then $x_H^{(l)*}>\theta_H$ and $x_M^{(l)*}<\theta_M$, so $\max_ix_i^{(l)*}-\min_ix_i^{(l)*}\ge x_H^{(l)*}-x_M^{(l)*}>\theta_H-\theta_M$. \qed

\subsection{Proof of Proposition~\ref{prop:signature}}
Both identities are read off \eqref{eq:xstar}; the ratio is well defined since $\theta_i\ne y^*$ and $1+\omega_{i,m}/c_{i,m}>0$. \qed

\subsection{Proof of Theorem~\ref{thm:dilution}}
Under a common stationary distribution, $\omega_{i,l}=\alpha_l\pi_i$ and $\bar\theta=\pi^\top\theta$ does not depend on $\alpha$.

(a) With $c_{i,l}=c_i$, $K_i(\alpha)=\pi_i^2\sum_l\alpha_l^2/c_i=s(\alpha)\pi_i^2/c_i$. Substituting into \eqref{eq:distortion} gives the displayed expression, and $s\mapsto sD/(1+sK^{(1)})$ has derivative $D/(1+sK^{(1)})^2$, so its absolute value is strictly increasing in $s$ when $D\ne0$. On the simplex $s(\alpha)\in[1/L,1]$, with the minimum at uniform $\alpha$ and the maximum at the vertices.

(b) Write $K_i^0$ for manipulability under the original weights and $D_0=\sum_iK_i^0(\theta_i-\bar\theta)$, $K_0=\sum_iK_i^0$. After adding the layer, $K_i(\varepsilon)=(1-\varepsilon)^2K_i^0+\varepsilon^2\pi_i^2/c_{i,L+1}$, so $D(\varepsilon)=(1-\varepsilon)^2D_0+\varepsilon^2V_{L+1}$ and $K(\varepsilon)=(1-\varepsilon)^2K_0+\varepsilon^2S_{L+1}$. Then $\PoM(\varepsilon)=|D(\varepsilon)|/(1+K(\varepsilon))$ and, since $D_0\neq0$, for small $\varepsilon$ the sign of $D(\varepsilon)$ equals that of $D_0$ and
\[
\frac{d}{d\varepsilon}\log\PoM(\varepsilon)\Big|_{0}=\frac{D'(0)}{D_0}-\frac{K'(0)}{1+K_0}=-2+\frac{2K_0}{1+K_0}=-\frac{2}{1+K_0}.
\]
At $\varepsilon=1$, $\PoM(1)=|V_{L+1}|/(1+S_{L+1})$, which gives the last claim. \qed

\subsection{Proof of Theorem~\ref{thm:linking}}
For a linked profile, agent $i$'s cost is $\tfrac12(\sum_lc_{i,l})(x_i-\theta_i)^2$ and their report enters $y$ with weight $\sum_l\omega_{i,l}=\pi_i$, so the argument of Theorem~\ref{thm:equilibrium} gives $K_i^{\mathrm{lk}}=\pi_i^2/\sum_lc_{i,l}$. By Cauchy-Schwarz, $(\sum_l\omega_{i,l})^2=(\sum_l\frac{\omega_{i,l}}{\sqrt{c_{i,l}}}\sqrt{c_{i,l}})^2\le\sum_l\frac{\omega_{i,l}^2}{c_{i,l}}\sum_lc_{i,l}$, with equality if and only if $\omega_{i,l}/\sqrt{c_{i,l}}$ is proportional to $\sqrt{c_{i,l}}$, that is, $\omega_{i,l}/c_{i,l}$ is constant in $l$. The gap formula is the Lagrange identity $(\sum_la_l^2)(\sum_lb_l^2)-(\sum_la_lb_l)^2=\sum_{l<m}(a_lb_m-a_mb_l)^2$ with $a_l=\omega_{i,l}/\sqrt{c_{i,l}}$, $b_l=\sqrt{c_{i,l}}$, divided by $\sum_lc_{i,l}$. \qed

\subsection{Proof of Theorem~\ref{thm:design}}
Under a common stationary distribution, $\sum_iK_i(\theta_i-\bar\theta)=\sum_l\alpha_l^2V_l$ and $K=\sum_l\alpha_l^2S_l$, so \eqref{eq:distortion} gives \eqref{eq:design}.

(a) If $V_l=0$ for some $l$, the vertex $e_l$ gives zero. If $V_l>0>V_m$, the function $t\mapsto t^2V_l+(1-t)^2V_m$ is continuous on $[0,1]$, positive at $t=1$ and negative at $t=0$, so it vanishes at some interior $t$, and the corresponding $\alpha$ gives zero distortion. Distortion is zero exactly when the numerator vanishes.

(b) Let $N(\alpha)=\sum_l\alpha_l^2V_l>0$, $Q(\alpha)=1+\sum_l\alpha_l^2S_l$, $f(\alpha)=N/Q$, and $f^*=\min_\Delta f$, attained since $\Delta$ is compact. For every $\alpha$, $f(\alpha)\ge f^*$ is equivalent to $N(\alpha)-f^*Q(\alpha)\ge0$, with equality if and only if $\alpha$ minimizes $f$. Hence the minimizers of $f$ are exactly the minimizers of $\phi(\alpha)=N(\alpha)-f^*Q(\alpha)=\sum_l\alpha_l^2(V_l-f^*S_l)-f^*$, and $\min_\Delta\phi=0$. Since $f^*\le f(e_l)=V_l/(1+S_l)<V_l/S_l$ for every $l$, all coefficients $a_l=V_l-f^*S_l$ are positive, so $\phi$ is strictly convex and has a unique minimizer on $\Delta$; by the first-order conditions it is interior with $\alpha_l\propto1/a_l$, which is \eqref{eq:alphastar}, and $\min_\Delta\sum_la_l\alpha_l^2=1/\sum_la_l^{-1}$. The condition $\min\phi=0$ therefore reads $1/\sum_l(V_l-f^*S_l)^{-1}=f^*$, which is \eqref{eq:root}. The left side of \eqref{eq:root} is continuous and strictly increasing in $f$ on $(0,\min_lV_l/S_l)$, equals $0$ at $f=0$ and tends to $+\infty$ at the right endpoint, so the root is unique. The limit claim follows by setting $S_l=0$ in \eqref{eq:alphastar}. \qed

\subsection{Proof of Theorem~\ref{thm:gradient}}
\emph{Existence of $Z$.} If $(I-C+\mathbf 1\pi^\top)v=0$, left-multiplying by $\pi^\top$ and using $\pi^\top(I-C)=0$, $\pi^\top\mathbf 1=1$ gives $\pi^\top v=0$, hence $(I-C)v=0$, so $v$ is a multiple of $\mathbf 1$ (the eigenvalue $1$ of a primitive stochastic matrix is simple), and $\pi^\top v=0$ forces $v=0$.

\emph{Derivative of $\pi$.} Since $C(\alpha)$ is primitive on the interior of the simplex, $\pi(\alpha)$ is the unique solution of $\pi^\top(I-C(\alpha))=0$, $\pi^\top\mathbf 1=1$, and by the implicit function theorem it is differentiable in $\alpha$. Differentiating $\pi^\top C=\pi^\top$ along $d$ gives $\dot\pi^\top(I-C)=\pi^\top\dot C$, and differentiating $\pi^\top\mathbf 1=1$ gives $\dot\pi^\top\mathbf 1=0$. Hence $\dot\pi^\top(I-C+\mathbf 1\pi^\top)=\pi^\top\dot C$, which is the first identity in \eqref{eq:pidot}. The remaining identities in \eqref{eq:pidot} and \eqref{eq:Ndot} are the chain rule applied to $\omega_{i,l}=\alpha_l(\pi^\top W^{(l)})_i$, $K_i=\sum_l\omega_{i,l}^2/c_{i,l}$, $N=\sum_iK_i(\theta_i-\pi^\top\theta)$, and $Q=1+\sum_iK_i$, using $\sum_i K_i\,\dot{\bar\theta}=K\dot\pi^\top\theta$; and $\log\PoM=\log|N|-\log Q$ with $N\ne0$.

\emph{Marginal layer.} Adding layer $L+1$ at weight $\varepsilon$ with the others rescaled is the direction $d=e_{L+1}-\alpha$ at the point $(\alpha,0)$, so $\dot C=W^{(L+1)}-C(\alpha)$. For an existing layer, $\dot\omega_{i,l}=-\omega_{i,l}+\alpha_l(\dot\pi^\top W^{(l)})_i$, so $\dot K_i=-2K_i+\rho_i$; for the new layer, $\omega_{i,L+1}=\varepsilon(\pi^\top W^{(L+1)})_i$ contributes $O(\varepsilon^2)$ to $K_i$ and nothing to $\dot K_i$ at $\varepsilon=0$. Substituting, $\dot N=-2N+\sum_i\rho_i(\theta_i-\bar\theta)-K\dot\pi^\top\theta$ and $\dot Q=-2K+\sum_i\rho_i$, so
\[
\frac{\dot N}{N}-\frac{\dot Q}{Q}=-2+\frac{2K}{1+K}+\frac{\sum_i\rho_i(\theta_i-\bar\theta)-K\dot\pi^\top\theta}{N}-\frac{\sum_i\rho_i}{1+K},
\]
which is \eqref{eq:marginal}. If $\dot\pi=0$ then every $\rho_i=0$ and $\mathcal R=0$. \qed

\subsection{Proof of Theorem~\ref{thm:algebraic}}
(a) Let $A(\alpha)=I-C(\alpha)$, an affine function of $\alpha$. Each entry of the adjugate $\operatorname{adj}A(\alpha)$ is an $(N-1)\times(N-1)$ minor of $A(\alpha)$, hence a polynomial in $\alpha$ of degree at most $N-1$. Since $C(\alpha)$ is primitive, $A(\alpha)$ has rank exactly $N-1$, so $\operatorname{adj}A(\alpha)\ne0$ has rank one; from $A\operatorname{adj}A=\operatorname{adj}A\cdot A=\det A\cdot I=0$, its columns lie in $\ker A=\operatorname{span}\{\mathbf 1\}$ and its rows in the left kernel $\operatorname{span}\{\pi^\top\}$, so $\operatorname{adj}A(\alpha)=\kappa(\alpha)\mathbf 1\pi(\alpha)^\top$ with $\kappa(\alpha)\ne0$. The function $\kappa=\operatorname{tr}\operatorname{adj}A$ is continuous and nonvanishing on the simplex (all layers primitive makes $C(\alpha)$ primitive at the vertices too), so it has constant sign; replacing $\operatorname{adj}A$ by its negative if necessary, take $a(\alpha)^\top$ to be the first row of $\operatorname{adj}A(\alpha)$, so $a_i=\kappa\pi_i$, $s=\sum_ja_j=\kappa>0$, and $\pi_i=a_i/s$. Then $(\pi^\top W^{(l)})_i=b_{i,l}/s$ with $b_{i,l}=\sum_ja_jW^{(l)}_{ji}$ of degree at most $N-1$; $\omega_{i,l}=\alpha_lb_{i,l}/s$; $K_i=\sum_l\alpha_l^2b_{i,l}^2/(c_{i,l}s^2)$ has numerator of degree at most $2N$; $\theta_i-\bar\theta=(\theta_is-a^\top\theta)/s$ has numerator of degree at most $N-1$. Hence $N(\alpha)=\tilde P/s^3$ with $\deg\tilde P\le3N-1$ and $Q(\alpha)=(s^2+\sum_{i,l}\alpha_l^2b_{i,l}^2/c_{i,l})/s^2=\tilde R/s^2$ with $\deg\tilde R\le2N$, so $\PoM=|\tilde P|/(s\tilde R)$; set $P=\tilde P$ and $R=s\tilde R$, of degree at most $3N-1$ and positive on the simplex.

(b) $\PoM=0$ if and only if $P=0$ since $R>0$. At the vertex $e_l$, $C=W^{(l)}$, $\pi=\pi^{(l)}$, $\omega_{i,l}=\pi^{(l)}_i$ and $\omega_{i,m}=0$ for $m\ne l$, so $N(e_l)=V^{(l)}$ and $P(e_l)=s(e_l)^3V^{(l)}$ has the sign of $V^{(l)}$. If two vertices have opposite signs, $P$ vanishes on the segment joining them by continuity; if some $V^{(l)}=0$, the vertex itself is a zero.

(c) On $[0,1]$ the function $\varepsilon\mapsto|P(\varepsilon)|/R(\varepsilon)$ is continuous, so a minimizer exists. If it is interior and $P$ does not vanish there, then $P/R$ is smooth in a neighborhood with the sign of $P$ constant, so the minimizer is a stationary point of $P/R$, that is a zero of $P'R-PR'$, whose degree is at most $(3N-2)+(3N-1)=6N-3$. Otherwise it is an endpoint or a zero of $P$. \qed

\end{document}